\documentclass[aps,prd,preprint,superscriptaddress,tightenlines,floatfix,showpacs,11pt]{revtex4-1}
\usepackage{graphics}
\usepackage{graphicx}
\usepackage{amsmath,amssymb}
\usepackage{tabularx}
\usepackage{array}
\usepackage[english,activeacute] {babel}

\begin{document}

\title{Shadow of a Renormalization Group Improved Rotating Black Hole}

\author{Luis A. S\'anchez}\email{lasanche@unal.edu.co}

\affiliation{Departamento de F\'\i sica, Universidad Nacional de Colombia,
A.A. 3840, Medell\'\i n, Colombia}

\begin{abstract}
We present a study on quantum gravity effects on the shadow of a rotating black hole (BH) obtained in the setting of the asymptotically safe gravity. The rotating metric, which results from a static regular one recently presented in the literature, is generated by using the generalized Newman-Janis algorithm. The novelty of the static regular metric lies in the fact that it is the outcome of an effective Lagrangian which describes dust whose spherically symmetric collapse is non-singular as a consequence of the antiscreening character of gravity at small distances. The effective Lagrangian includes a multiplicative coupling, denoted as $\chi$, with the Lagrangian of the collapsing fluid. The resulting exterior metric for large radii depends on a free parameter $\xi$ which captures the quantum gravity effects. The form of the coupling $\chi$ and its connection with the quantum parameter $\xi$ are determined by the running of the Newton coupling $G(k)$ along a renormalization group trajectory that stops at the ultraviolet non-gaussian fixed point of the asymptotic safety theory for quantum gravity. Varying both the spin parameter $a_{\star}$ and the quantum parameter $\xi$, we explore the quantum gravity effects on several astronomical observables used to describe the morphology of the shadow cast by rotating BHs. In order to obtain constraints on the parameter $\xi$, we confront our results with the recent Event Horizon Telescope (EHT) observations of the shadows of the supermassive BHs M87$^*$ and Sgr A$^*$. We find that the ranges of variation of all the studied shadow observables fall entirely within the ranges determined by the EHT collaboration. We then conclude that the current astronomical data do not rule out the renormalization group improved rotating BH.\\
 
\noindent
{\bf keywords}: Quantum aspects of black holes, asymptotic safety, black hole shadow.
\end{abstract}


\maketitle

\section{\label{sec:intr}Introduction}
It is well known that, in the classical theory of GR, singularity theorems establish that, as long as certain energy conditions are satisfied, the gravitational collapse of a massive body inevitably gives rise to a curvature singularity that renders geodesic incomplete the space-time geometry \cite{r1,r2,r3}. Moreover, the cosmic censorship hypothesis affirms, in its weak form, that the singularity arising from the gravitational collapse to a BH is concealed by an event horizon which hidden it from detection by any external observer \cite{r4}. The strong version positions this conjecture at the level of a universal principle as asserts that {\it every} singularity arising from gravitational collapse is {\it always} hidden behind an event horizon. So, the strong version excludes the physical existence of naked singularities. Since gravitational singularities (or singular space-times) imply the loss of predictability of the classical GR, they are considered as unphysical pathologies in the ultraviolet (UV) regime where the theory breaks down. Thus, there is a general consensus that the solution to the so-called ``singularity problem'' requires a quantum theory of gravity that, at trans-Planckian energies, should supersede classical GR and that should provide us with non-singular BH solutions (see Ref.~\cite{r5} for a recent review on regular BH).\\

\noindent
Among the several proposals for a quantum theory of gravity, asymptotic safety (AS) is a promising contender that, using the methods of the standard quantum field theory and the techniques of the functional renormalization group (FRG), gives rise to a non-perturbative renormalizable quantum theory of the gravitational interaction \cite{r6}. The basic ingredient of the theory is the conjecture of the existence of an UV non-gaussian fixed point which attracts the flow of the scale dependent quantum effective action for gravity. The fixed point determines a finite transition scale, identified as the Planck scale, beyond which quantum scale invariance is realized such that a finite number of relevant dimensionless couplings, as the Newton coupling $G(k)k^2$, become constant without dependence of the energy scale. In contrast, the corresponding dimensionfull Newton coupling $G(k)$ falls down running as the inverse squared of the energy scale and, consequently, quantum fluctuations weaken the gravitational interaction in the UV \cite{r7,r8,r9}. It is just this antiscreening effect of the quantum fluctuations at high energies that opens the possibility of solving the singularity trouble in the AS scenario. The running Newton coupling leads to the construction of the so-called renormalization group improved (RGI) metrics which allow including the leading quantum gravity effects on gravitational phenomena in the strong field regime \cite{r10,r11,r12,r13,r14}. In this regard, the problem of the singularity-free gravitational collapse in AS was studied for the first time in \cite{r15} where it was concluded that stronger deviations from GR than the provided by AS, are required to prevent the final singularity. In \cite{r16}, by analyzing the modifications induced by AS gravity on the gravitational collapse of generic Vaidya spacetimes, Misner-Sharp mass functions were obtained that lead to infer a softening of the singularity as compared with GR. The problem was revisited in \cite{r17} where, by assuming a RGI regular exterior metric, the effective interior RGI metric is deduced from compatibility with the exterior one, resulting in an regular interior dust collapse. Indications that the weakening of gravity in the UV in the context of the AS theory could solve the singularity problem, have also been discussed in \cite{r10,r18,r19,r20}.\\

\noindent
Recently, a more satisfactory top-down approach to the singularity resolution has been adopted by constructing an effective Lagrangian for the gravitational collapse of dust that includes a multiplicative coupling $\chi$ with the matter Lagrangian \cite{r21}. The form of this coupling is obtained from the running of the Newton coupling $G(k)$ along a renormalization group trajectory that ends at the Reuter ultraviolet (UV) fixed point of the AS theory. In the resulting exterior metric, the coupling $G(k)$ gives rise to a Misner-Sharp mass function ${\cal M}(r,\xi)$ depending on a free parameter $\xi$ linked to $\chi$ by suitable junction conditions. The exterior metric remains everywhere regular without the existence of singularities and, for large radii, the associated RGI static metric is a Schwarzschild-like one from which the classical spherically symmetric line element is recovered when $\xi=0$. Since there is growing observational support that astrophysical BH candidates rotate, it is pertinent to construct the RGI rotating counterpart of the static metric and verify if the RGI rotating version remains regular. For this purpose we generate the rotating metric through the Newman-Janis (NJ) algorithm \cite{r22} and, in the light of a theorem that establish the necessary and sufficient conditions for the absence of scalar curvature singularities \cite{r23}, we discuss the issue of the regularity of the RGI rotating BH.\\

\noindent
On the other hand, the recent observation of the shadow images of the supermassive BHs M87$^*$ and Sgr A$^*$ by the Event Horizon Telescope (EHT) collaboration \cite{r24,r25}, has opened new possibilities to investigate gravity theories in the strong field regime. The important fact about these images is that they are determined only by the space-time background metric making visible horizon scale features of the BHs. Immediately after the publication of the EHT results, a considerable number of interesting papers, focused on testing GR, alternative theories of gravity as well as quantum gravity scenarios, have appeared in the literature (see e.g. \cite{r26} and references therein for a large and, already, incomplete list of works). In the context of the AS theory, investigations have been done to identify quantum gravity modifications to the shadows of both non-rotating and rotating regular RGI BHs, and to constraint the corresponding quantum parameter \cite{r27,r28,r29,r30}.\\

\noindent
In this paper we aim to study quantum gravity effects on the shadow cast by a BH whose specetime is described by the RGI rotating metric that, after using the NJ algorithm, results from the static exterior RGI metric obtained in \cite{r21}. In this context, we analyze the quantum gravity effects on several selected astronomical observables used to describe the shadow shape. By confronting our results with the recent EHT observations of the shadows of the supermassive BHs M87$^*$ and Sgr A$^*$, we also search for constraints on the quantum parameter $\xi$ appearing in the mass function ${\cal M}(r,\xi)$.\\

\noindent
This paper is organized as follows. In Sect. 2 we present a sketch of the procedure in \cite{r21} to obtain the exterior metric for a static RGI regular BH from the collapse of dust in the AS scenario. In Sect. 3 we use the NJ algorithm to construct the metric of the RGI rotating BH and we discuss whether this metric is regular or not. In Sect. 4 we study the photon geodesics, the general formulae describing the shadow contour of a rotating BH whose metric involves the mass function ${\cal M}(r,\xi)$, and analyze the shadow of the RGI rotating BH. In Sect.5 we describe the shadow observables that we will use, confront our results with the EHT data, and search for constraints on the parameter $\xi$. In the last section we present our conclusions.
%
\section{\label{sec:sec2}The static regular black hole in asymptotic safety}
In this section we briefly outline the procedure to obtain the static regular BH in AS from the collapse of dust (for details see Ref.~\cite{r21}). The starting point is the introduction of an action for an effective lagrangian which includes a gravity-matter coupling $\chi(\cal E)$ with the lagrangian ${\cal L}_{m}=-{\cal E}$ of a non-dissipative fluid with rest-mass density $\rho$ and four-velocity $u^\mu$ satisfying $\delta \rho/\rho=\delta {\cal E}/(p({\cal E})+{\cal E})$ and $u_\mu u^\mu=-1$. This action reads
\begin{equation}\label{eq1}
S = \frac{1}{16\pi G_0}\int d^4x\sqrt{-g}[R+2\chi({\cal E}){\cal L}_{m}],
\end{equation}
where $G_0$ is the Newton coupling and $R$ is the Ricci scalar. The field equations that follow from the total variation of $S$ are
\begin{equation}\label{eq2}
R_{\mu \nu}-\frac{1}{2}g_{\mu \nu}R = 8\pi G({\cal E})T_{\mu \nu}-\Lambda({\cal E})g_{\mu \nu} \equiv T^{\rm eff(-)}_{\mu \nu},
\end{equation}
where $8\pi G({\cal E})=\partial ({\cal E}\chi)/\partial {\cal E}$, $\Lambda({\cal E})=-{\cal E}^2\partial \chi/\partial {\cal E}$, and $T_{\mu \nu}=(p({\cal E})+{\cal E})u_\mu u_\nu +pg_{\mu \nu}$, with $G({\cal E})$ and $\Lambda({\cal E})$ being the effective Newton coupling and the effective cosmological constant, respectively. Now, for a spherically collapsing fluid, the metric in co-moving coordinates is
\begin{equation}\label{eq3}
ds^2_{(-)}=g_{00}dt^2 +g_{11}dr^2+g_{22}d\Omega^2=-e^{-2\nu(r,t)}dt^2 +e^{-2\psi(r,t)}dr^2+C^2(r,t)d\Omega^2.
\end{equation}
where $d\Omega^2$ is the metric on the unit 2-sphere. Then, the field equations acquire the form
\begin{eqnarray}\label{eq4}
\frac{1}{C^2}\frac{d{\cal M}_{\rm eff(-)}}{dr}\left(\frac{dC}{dr}\right)^{-1}&=&8\pi G({\cal E}){\cal E}+\Lambda({\cal E})
= {\cal E}\chi({\cal E}) \equiv {\cal E}_{\rm eff(-)},\\ \label{eq5}
\frac{1}{C^2}\frac{d{\cal M}_{\rm eff(-)}}{dt}\left(\frac{dC}{dt}\right)^{-1}&=&-8\pi G({\cal E}){\cal E}+\Lambda({\cal E})
\equiv -p_{\rm eff(-)},\\ \label{eq6}
\frac{d}{dr}\left(\frac{dC}{dt}\right)&=& \frac{dC}{dt}\frac{d\nu}{dr}+\frac{dC}{dr}\frac{d\psi}{dt},
\end{eqnarray}
where ${\cal M}_{\rm eff(-)}$ is the interior Misner-Sharp mass function given by
\begin{equation}\label{eq7}
{\cal M}_{\rm eff(-)}=C\left(1-\left(\frac{dC}{dr}\right)^2 e^{-2\psi}+\left(\frac{dC}{dt}\right)^2 e^{-2\nu}\right),
\end{equation}

\noindent
After using the Bianchi identity to integrate Eq.~(\ref{eq6}), the interior metric (\ref{eq3}) can be written as
\begin{equation}\label{eq8}
ds^2_{(-)}=-dt^2 +\frac{1}{1-K r^2}\left(\frac{dC}{dr}\right)^2dr^2+C^2 d\Omega^2,
\end{equation}
where the integration constant $K$ is the curvature of the 3-space. Using the dimensionless scale factor $a(t)$ to rescaling the area-radius $C$ as $C=a(t)r$ and rescaling the Misner-Sharp function as ${\cal M}_{\rm eff(-)}=r^3 m_{\rm eff(-)}$, Eqs.~(\ref{eq4}) and (\ref{eq7}) produce
\begin{eqnarray}\nonumber
m_{\rm eff(-)}(a)&=&\frac{a^3}{3}{\cal E}_{\rm eff(-)}=a\left(\left(\frac{da}{dt}\right)^2+K\right) \\ \label{eq9}
&=&-aV(a)=-a\left(-\frac{a^2}{3}\int^{{\cal E}(a)}_0G({\cal E}^{\prime})d{\cal E}^{\prime}\right).
\end{eqnarray}

\noindent
At this point, the next key step to connect with the AS gravity is to assume that the evolution of $G(k)$ with the energy scale $k$ follows a renormalization group trajectory that ends at the UV fixed point $g_{\star}$ and, accordingly, is given by \cite{r31}
\begin{equation}\label{eq10}
G(k) = \frac{G_0}{1+G_0 k^2/g_{\star}},
\end{equation}
where $g_{\star}=570\pi/833$. Here it must be remarked that it is also assumed that the presence of the collapsing fluid does not perturb the flow enough as to alter the fixed point. That this assumption is robust, it has been shown in \cite{r32}.\\

\noindent
If the collapsing fluid is dust, then $p=0$, ${\cal E}\propto a^{-3}$ and $\rho={\cal E}$. Additionally, the link between ${\cal E}$ and $k$ is achieved by using the identification between the proper distance $d$ and the energy scale $k$ in AS gravity, i.e., $d\sim 1/k$ which produces $d\sim {\cal E}^{-1/2}$, and the running Newton coupling becomes
\begin{equation}\label{eq11}
G({\cal E}) = \frac{G_0}{1+\xi {\cal E}},
\end{equation}
where the parameter $\xi$, with units of inverse mass to the fourth power, includes the constants $G_0$ and $g_{\star}$ but it is a free parameter of the AS theory that must be constrained from the observational data. The link implies
\begin{equation}\label{eq12}
\chi({\cal E})=\frac{{\rm Log}(1+\xi{\cal E})}{\xi{\cal E}}, \quad \quad \Lambda({\cal E})=\frac{{\rm Log}(1+\xi{\cal E})}{\xi}-\frac{{\cal E}}{1+\xi{\cal E}},
\end{equation}

\noindent
When the present formalism is brought to the Oppenheimer-Snyder approach to the gravitational collapse of dust \cite{r33}, it turns out that $m_{\rm eff}\rightarrow m_0$, $T_{\mu \nu}={\cal E}u_{\mu}u_{\nu}$ and ${\cal E}=3m_0/a^3$,  and the scale factor obeys the equation
\begin{equation}\label{eq13}
\frac{da}{dt}=-\sqrt{\frac{{\rm Log}(1+3m_0\xi/a^3)}{3\xi}a^2-K}.
\end{equation} 
In the limit $t\rightarrow \infty$, $a(t)$ behaves as
\begin{equation}\label{eq14}
a \sim e^{-t^2/4\xi},
\end{equation}
such that, provided $\xi>0$, the value $a=0$ is never reached. This means that, due to the antiscreeneing character of gravity at small distances in AS, a collapsing boundary $C(t,r_b)=a(t)r_b$ at an interior co-moving boundary $r=r_b$, never collapses to a singularity. The space-time is then geodesically complete.\\

\noindent
Following \cite{r34,r35,r36}, the matching with the exterior geometry proceeds as follows. First, a generic exterior spherically symmetric line element is assumed
\begin{equation}\label{eq15}
ds^2_{(+)}=-f(R)dT^2+f(R)^{-1}dR^2+R^2 d\Omega^2
\end{equation}
with $f(R)=1-2{\cal M}(R)/R$ and where now $\lbrace T,R,\theta,\phi\rbrace$ are exterior coordinates. Now, for the collapsing boundary $C(t,r_b)=a(t)r_b$, the induced metric on the matching surface $\Sigma$ is written, in co-moving coordinates, as 
\begin{equation}\label{eq16}
ds^2_{\Sigma}=-dt^2+a^2 r^2_b d\Omega^2
\end{equation}
Parameterizing the collapsing boundary as $R=R_b(T)$, Eq.~(\ref{eq16}) takes the form
\begin{equation}\label{eq17}
ds^2_{\Sigma}=-\left(f(R_b)-\frac{1}{f(R_b)}\left(\frac{dR_b}{dT}\right)^2\right)dT^2+a^2 R^2_b d\Omega^2.
\end{equation}

\noindent
For the interior metric, the second fundamental form (which reflects the extrinsic geometry of the interior surface), in co-moving coordinates, is
\begin{equation}\label{eq18}
K^{(-)}_{tt}=0, \quad \quad   K^{(-)}_{\theta \theta}=ar_b\sqrt{1-Kr^2_b},
\end{equation}
and the exterior extrinsic curvature is given by
\begin{equation}\label{eq19}
K^{(+)}_{tt}=-\frac{1}{2\Delta(R_b)}\left(2\frac{d^2 R_b}{dt^2}+\frac{df(R_b)}{dR}\right), \quad \quad   K^{(+)}_{\theta \theta}
=R_b\Delta(R_b),
\end{equation}
where
\begin{equation}\label{eq20} 
\Delta(R_b)=\sqrt{1-2\frac{{\cal M}(R_b)}{R_b} + \left(\frac{dR_b}{dt}\right)^2}.
\end{equation}
Finally, imposing the junction conditions $R_b(T(t))=a(t)r_b$ and
\begin{equation}\label{eq21}
K^{(+)}_{tt}=K^{(-)}_{tt}, \quad \quad K^{(+)}_{\theta \theta}=K^{(-)}_{\theta \theta},
\end{equation}
the functional form of the exterior Misner-Sharp mass ${\cal M}(R,\xi)$ is obtained as
\begin{equation}\label{eq22}
{\cal M}(R,\xi)=\frac{R^3}{6\xi}{\rm Log}\left(1+\frac{6\xi M_0}{R^3}\right)
\end{equation}
with $M_0=m_0r^3_b/2$. Then, for $R<<1$
\begin{equation}\label{eq23}
{\cal M}(R,\xi)=\frac{R^3}{6\xi}{\rm Log}\left(\frac{6M_0\xi}{R^3}\right)+O(R^6),
\end{equation}
and given that $R\geq R_b=a(t)r_b$ and $a(t)>0$, the solution (\ref{eq22}) is regular. Note that again $\xi>0$ is required in order for ${\cal M}(R)$ to be well defined in this limit. On the other hand, in the IR limit $R>>1$ ($\xi<<1$)
\begin{equation}\label{eq24}
{\cal M}(R,\xi)=M_0-\frac{3M^2_0\xi}{R^3}+O(\xi^2),
\end{equation}
and we see that for $\xi=0$ the classical constant Schwarzschild BH mass $M_0$ is recovered. It is then clear that the free parameter $\xi$ captures the quantum gravity effects of the AS scenario.
%
\section{\label{sec:sec3}The metric of the RGI rotating black hole}
In the previous section we saw how the matching conditions imposed between the interior and the exterior geometry, make the running Newton coupling $G(k)$ of the AS theory turns into the Misner-Sharp mass function ${\cal M}(R,\xi)$ appearing in the Schwarzschild-like form of the exterior metric. Renaming the exterior coordinates $\lbrace T,R,\theta,\phi \rbrace$ to the most commonly used $\lbrace t,r,\theta,\phi \rbrace$, the exterior $ds^2_{(+)}\equiv ds^2$ static regular metric is given by
\begin{equation}\label{eq25}
ds^2=-\left(1-\frac{2{\cal M}(r,\xi)}{r}\right)dt^2+\left(1-\frac{2{\cal M}(r,\xi)}{r}\right)^{-1}dr^2+r^2 
d\Omega^2.
\end{equation}
The associated rotating BH is obtained from the generalized NJ algorithm \cite{r22}, which is a procedure that includes the following steps (for details see \cite{r37,r38,r39,r40}). In the first step, a general static spherically symmetric metric is considered
\begin{equation}
ds^{2}=-f(r)dt^{2}+g(r)^{-1}dr^{2}+h(r)d\Omega^{2}.  \label{eq26}
\end{equation}
and the Boyer-Lindquist coordinates ($t,r,\theta,\phi$) are transformed into advanced null coordinates ($u,r,\theta,\phi$). Then, the contravariant form of the line element is written in terms of a null tetrad $Z^\mu_a$ and, to include rotation, the advanced null coordinates $x^\mu$ are transformed into a new set of complex coordinates ${\tilde x}^\mu = x^{\mu} + ia (\delta_r^{\mu} - \delta_u^{\mu})\cos\theta$, such that the metric coefficients take the new form: $f\left(r\right)\rightarrow F\left(r,a,\theta\right)$, $g\left(r\right)\rightarrow G\left(r,a,\theta\right)$, and $h\left(r\right)\rightarrow\Sigma=r^{2}+a^{2}\cos^{2}{\theta}$, and such that the null tetrad vectors $Z^\mu_a$ undergo the transformation $Z^\mu_a\rightarrow {\tilde Z}^\mu_a({\tilde x}^\nu)$ with the requirement that this transformation recovers the old tetrad and the old line element when ${\tilde x}^\mu = x^{\mu}$. In the last step, the new metric is reversed back to Boyer-Lindquist coordinates. In this way, the rotating counterpart of the metric in Eq.~(\ref{eq26}) takes the form
\begin{equation}\label{eq27}
\begin{split}
ds^{2}=&-\frac{\Sigma\left(g(r)h(r)+a^{2}\cos^{2}{\theta}\right)}{\left(k(r)+a^{2}\cos^{2}{\theta}\right)^{2}}dt^{2}+\frac{\Sigma}{g(r)h(r)+a^{2}}dr^{2}-2a\sin^{2}\theta\Sigma\left[\frac{k(r)-g(r)h(r)}{\left(k(r)+a^{2}\cos^{2}{\theta}\right)^{2}}\right] d\phi dt\\
&+\Sigma d\theta^{2}+\Sigma\sin^{2}\theta\left[1+a^{2}\sin^{2}\theta\frac{2k(r)-g(r)h(r)+a^{2}\cos^{2}\theta}{\left(k(r)+a^{2}\cos^{2}{\theta}\right)^{2}}\right]d\phi^{2}.
  \end{split}
\end{equation}
Comparing now Eqs.~(\ref{eq25}) and (\ref{eq26}) and replacing into Eq.~(\ref{eq27}), the RGI rotating metric acquires the form
\begin{eqnarray}\nonumber
ds^2&=&-\Big(1-\frac{2{\cal M}(r,\xi) r}{\Sigma}\Big)dt^2-\frac{4a{\cal M}(r,\xi) r\sin^2\theta}{\Sigma}dtd\phi+\frac{\Sigma}{\Delta}dr^2+\Sigma{d\theta^2}\\\label{eq28}
&+&\sin^2\theta\bigg[r^2+a^2+\frac{2a^2 {\cal M}(r,\xi) r}{\Sigma}\sin^2\theta\bigg]d\phi^2,
\end{eqnarray}
where $a$ is interpreted as the BH spin and
\begin{equation} 
\Delta=r^2-2{\cal M}(r,\xi)r+a^2,
\label{eq29}
\end{equation}
\begin{equation} 
\Sigma=r^2+a^2 \cos{\theta}.
\label{eq30}
\end{equation}
Note that in the limit $\xi<<1$, ${\cal M}(r,\xi)$ is given by Eq.~(\ref{eq24}) from which we retrieve the classical expression for $\Delta$ when $\xi=0$.\\

\noindent
Let us now discuss the regularity of the RGI rotating metric in Eq.~(\ref{eq28}). This metric is Petrov type $D$ and Segrev type $[(1,1)(11)]$ with an algebraically complete set of four second order invariants \cite{r37}. For this type of metrics, it has been proved in \cite{r23} that if ${\cal M}(r,\xi)$ is a $C^3$ function, all the second order scalar curvature invariants of the metric (\ref{eq28}) are finite at ($r=0$, $\theta=\pi/2$) {\it if, and only if}, ${\cal M}(0,\xi)={\cal M}^{\prime}(0,\xi)={\cal M}^{\prime \prime}(0,\xi)=0$ (here, the prime denotes derivative with respect to $r$). For the mass function ${\cal M}(r,\xi)$ in Eq.~(\ref{eq22}) is direct to show that: $\lim_{r\to 0}{\cal M}(r,\xi)=0$, $\lim_{r\to 0}{\cal M}^{\prime}(r,\xi)=0$, but $\lim_{r\to 0}{\cal M}^{\prime\prime}(r,\xi)\rightarrow \infty$. So, in the light of this theorem, the rotating metric does not satisfy the necessary and sufficient regularity conditions. Notice however that, as mentioned in the previous section, the collapsing dust ball never reaches the singularity and therefore, the theorem, which involves an evaluation at $r=0$, does not apply. In other words, as it happens for the static exterior metric, given that $R\geq R_b=a(t)r_b$ and $a(t)>0$, the rotating metric in Eq.~(\ref{eq28}) is regular, that is, this metric inherits the absence of singularities of the collapsing dust ball from which it comes.\\
\begin{figure}[t!]
 \includegraphics[width=.44\linewidth]{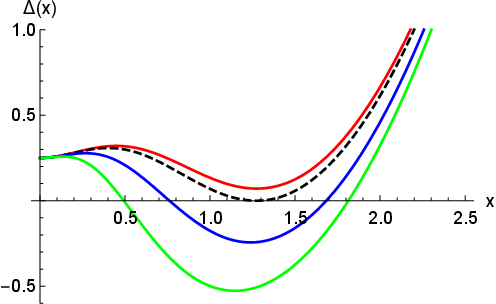}
  \includegraphics[width=.44\linewidth]{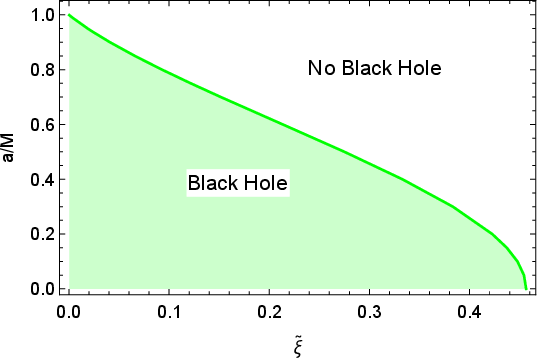}
\caption{Left panel: Plot of $\Delta(x)$ versus $x$, with $x=r/M_0$, for $a_{\star}=0.5$ and for $\tilde{\xi} =\tilde{\xi}^{(0.5)}_{c}=0.275119633455$ (black dashed), $\tilde{\xi} >\tilde{\xi}^{(0.5)}_{c}$ (red), $\tilde{\xi}=0.15 <\tilde{\xi}^{(0.5)}_{c}$ (blue), and $\tilde{\xi}=0.05 <\tilde{\xi}^{(0.5)}_{c}$ (green). Right panel: the parameter space $(a/M_0,\tilde{\xi})$. The green curve corresponds to the critical values $\tilde{\xi}_c$ for each value of $a_{\star}$.}
\label{F1}
\end{figure}
\begin{figure}[b!]
 \includegraphics[width=.48\linewidth]{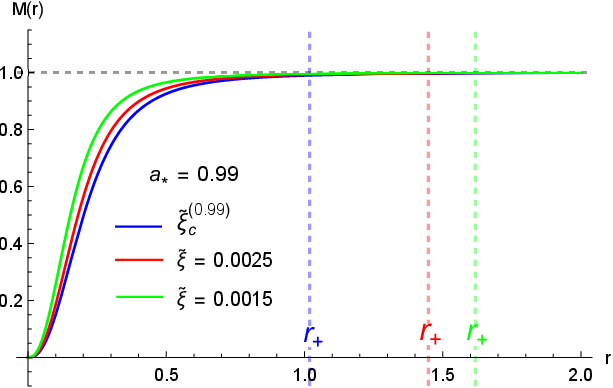}
\caption{Plots of the mass function ${\cal M}(r,\xi)$ for $a_{\star}=0.99$ and for the values of $\tilde{\xi}$ as shown. The vertical dashed lines locate the position of the outer event horizon $r_{+}$ for the same value of $a_{\star}$ and for each one of the values of $\tilde{\xi}$.}
\label{F2}
\end{figure}

\noindent
The BH horizons are the solutions to the transcendental equation $\Delta=0$. In the left panel of Fig.~\ref{F1} we plot $\Delta(x)$ versus $x$ where $x=r/M_0$. The plot shows that for fixed values of the spin parameter 
$a_{\star}=a/M_0$ and as long as the dimensionless parameter $\tilde{\xi}=\xi M_0^4$ is lees than a critical value $\tilde{\xi}_{c}$, there are two horizons: one inner Cauchy horizon $r_{-}$ and one outer event horizon $r_{+}$. When $\xi=\xi_{c}$ the two horizons merge, and for $\tilde{\xi}>\tilde{\xi}_{c}$ a scalar remnant appears. The right panel of Fig.~\ref{F1} shows the parameter space $(a/M_0,\tilde{\xi})$. The green curve corresponds to the critical values $\tilde{\xi}_c$ for each value of $a/M_0$ in such a way that for $\tilde{\xi} < \tilde{\xi}_c$ (shaded region) there are RGI rotating BHs, whilst for $\tilde{\xi} > \tilde{\xi}_c$ (unshaded region) one has a scalar remnant. Consistent with our discussion in the previous section, we will consider for our study values of $\tilde{\xi}$ satisfying $0<\tilde{\xi}\leq \tilde{\xi}_{c}$.\\

\noindent
At this point, it should be highlighted that the horizon becomes more compact as $\tilde{\xi}$ grows (see left panel of Fig.~\ref{F1}). Moreover, the line element in  Eq.~(\ref{eq28}) has the form of a Kerr-like metric with the BH mass replaced by the ``effective" mass  ${\cal M}(r,\xi)$ which, consequently, contains all the information on the quantum gravity effects on the space-time geometry. Fig.~\ref{F2} are plots of ${\cal M}(r,\xi)$ for growing values of $\tilde{\xi}$: $\tilde{\xi}=\tilde{\xi}_c^{(0.99)}=0.003425142584$ which is the critical value for $a/M_0=0.99$ (blue curve), $\tilde{\xi}=0.0025$ (red curve), and $\tilde{\xi}=0.0015$ (green curve). Since all of these values of $\tilde{\xi}$ are allowed for  $a/M_0=0.99$, the blue, red, and green vertical dashed lines locate the position of the outer event horizon $r_{+}$ for this value of $a/M_0$ and for the referred values of $\tilde{\xi}$, respectively. If we fix our attention on the horizon $r_{+}$ for $\tilde{\xi}=\tilde{\xi}_c^{(0.99)}$, we see that even tough the difference in the ``effective" mass hidden behind the horizon is negligible for different values of $\tilde{\xi}$, the quantum gravity effect on the location of the horizon is considerable. For example, when passing from $r_{+}$ for $\tilde{\xi}=0.0025$ to $r_{+}$ for 
$\tilde{\xi}=\tilde{\xi}_c^{(0.99)}$, the size of the horizon decreases by $\sim 30\%$. The shrinking of the horizon is then a quantum and strong-field effect and should gives rise to strong distortions of the shadow of the RGI rotating BH. As we will see in the next section, this is just what happens specially for high values of $a/M_0$ and 
$\tilde{\xi}$\\

\section{\label{sec:sec4}Photon geodesics and shadow}
The space-time with the metric in Eq.~(\ref{eq28}) is stationary and axisymmetric. There are, then, two Killing vector fields and, consequently, two conserved quantities. They are the energy $E$ and the axial component of the angular momentum $L_z$. Associated to the Killing-Yano tensor field, there is a third conserved quantity, the Carter constant ${\cal Q}$. These three quantities are related to each other through the quadratic integral of motion 
${\cal K}$ in the form \cite{r41}
\begin{equation}
{\cal Q} = {\cal K} - (L_z-aE)^2.
\label{eq31}
\end{equation}
The null-geodesic equations are
\begin{eqnarray}\label{eq32}
\Sigma \frac{\dot{r}}{E}&=& \pm \sqrt{{\cal R}(r)},\\ \label{eq33}
\Sigma \frac{\dot{\theta}}{E}&=& \pm \sqrt{\Theta(\theta)},\\  \label{eq34}
\Sigma \frac{\dot{\phi}}{E}&=& \frac{\eta^2}{\sin^2\theta}+\frac{a}{\Delta}\left[2r{\cal M}(r,\xi)-a\eta\right], \\ \label{eq35}
\Sigma \frac{\dot{t}}{E}&=& -a^2 \sin^2\theta+ \frac{1}{\Delta}[\left(r^2+a^2)^2-2ar{\cal M}(r,\xi)\eta\right],
\end{eqnarray}
where the dots denote derivative respect to an affine parameter along the geodesic and
\begin{eqnarray}\label{eq36}
{\cal R}(r)&=& r^4+\left(a^2-\eta^2-\chi\right)r^2+2r\left[\chi+(\eta-a)^2\right]{\cal M}(r,\xi)-a^2\chi,\\ \label{eq37}
\Theta(\theta)&=& \chi +a^2\cos^2\theta-\eta^2\cot^2\theta, 
\end{eqnarray}
and where the two conserved parameters $\chi$ and $\eta$ are defined as
\begin{equation}\label{eq38}
\chi = \frac{{\cal Q}}{E^2}, \quad \quad \eta = \frac{L_z}{E}.
\end{equation}

\noindent
As stated above, the shadow of a BH is a pure strong-field effect that probes gravity theories on horizon scales and, therefore, the background spacetime geometry. The boundary of the BH shadow corresponds to the innermost unstable spherical photon orbit with radius $r_p$ determined by the conditions ${\cal R}(r)=0$ and $\partial{\cal R}(r)/\partial r=0$, which provide the following expressions for the conserved impact parameters \cite{r42}
\begin{eqnarray}\label{eq39}
\chi &=& \frac{r^3}{a^2(r-{\cal M}-r{\cal M}^{\prime})^2}\left[4a^2{\cal M}-r(r-3{\cal M})^2-2r{\cal M}^{\prime}
(r^2-3r{\cal M}+2a^2)-r^3({\cal M}^{\prime})^2 \right], \\  \label{eq40}
\eta &=& \frac{(r^2-a^2){\cal M}-r\Delta -r(r^2+a^2){\cal M}^{\prime}}{a(r-{\cal M}-r{\cal M}^{\prime})},
\end{eqnarray}
where the prime denotes derivative with respect to $r$.\\

\noindent
Let us now consider photon sources located at infinity and distributed uniformly in all directions, and an observer also at infinity. The shadow of the BH is obtained by solving the scattering problem of photons falling onto the BH from infinity as seen by the observer. The celestial coordinates of the observer $(x,y)$ are the impact parameters of photons falling just into the innermost unstable photon orbit and correspond to the apparent angular distances of the image on the celestial sphere measured from the direction of the observer's line of sight. In this sense, $x$ and $y$ are measured in the directions perpendicular and parallel to the projected rotation axis onto the celestial sphere, respectively. Let $\iota$ be the inclination angle of the observer defined as the angle between its line of sight and the rotation axis of the BH. Then, with $(p^{(t)},p^{(r)},p^{(\theta)},p^{(\phi)})$ being the tetrad components of the photon momentum respect to a locally non-rotating reference frame, the coordinates $(x,y)$ can be written in terms of the parameters $(\chi, \eta)$ and the angle $\iota$ as
\begin{eqnarray}\label{eq41}
x &=& \lim_{r\to \infty}\frac{-r p^{(\phi)}}{p^{(t)}}=-\eta \csc \iota, \\ \label{eq42}
y &=& \lim_{r\to \infty}\frac{r p^{(\theta)}}{p^{(t)}}=\pm \sqrt{\chi+a^2\cos^2\iota-\eta^2\cot^2\iota},
\end{eqnarray}
where the plus sign is for prograde orbits and the minus sign is for retrograde orbits. The apparent shape of the rotating BH (its shadow) is obtained by doing the parametric plot of $y(r)$ versus $x(r)$ for fixed values of $M_0$, $a$, 
$\iota$ and $\xi$ (in the following we will write $M$ instead of $M_0$ for simplicity).\\
\subsection{\label{sec:sec4.1}The shadow of the RGI rotating black hole}
\begin{figure}[t!]
 \includegraphics[width=.48\linewidth]{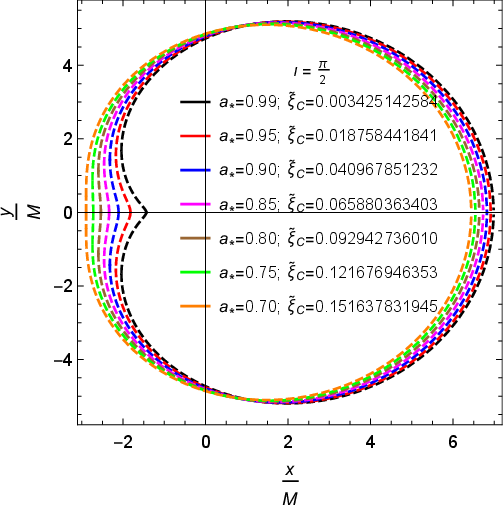}
\caption{Shadows cast for the RGI rotating BH for the observer inclination angle $\iota=\pi/2$. Each shadow silhouette is constructed for the values $(a_{\star}, \tilde{\xi}^{(a_{\star})}_c)$ as shown. The left cusp effect fades as $a_{\star}$ decreases.}
\label{F3}
\end{figure}
\begin{figure}[h!]
 \includegraphics[width=.32\linewidth]{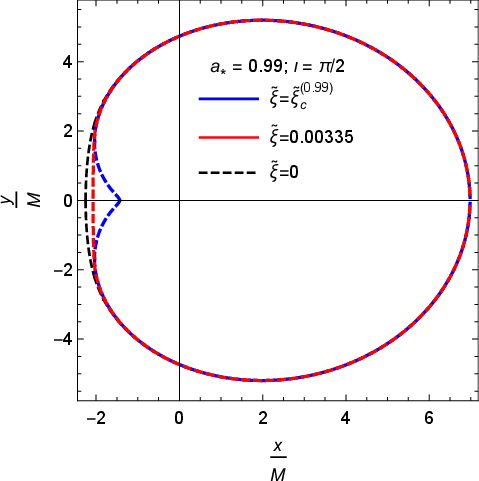}
 \includegraphics[width=.32\linewidth]{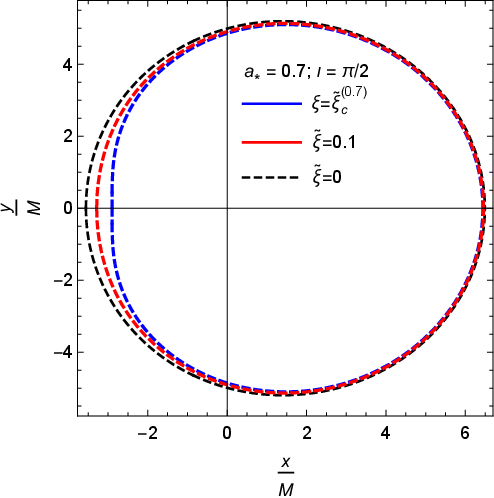}
 \includegraphics[width=.32\linewidth]{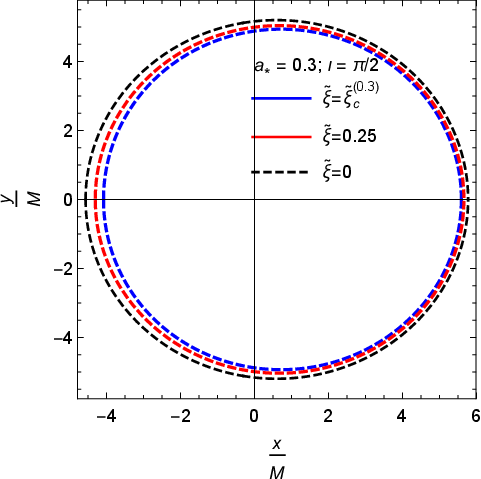}
\caption{The same as in Fig.~\ref{F3} but including values of $\tilde{\xi}$ less than $\tilde{\xi}_c$ and for $a_{\star}=0.99$ (left panel), $a_{\star}=0.7$ (central panel), and $a_{\star}=0.3$ (right panel). For each value of $a_{\star}$, the comparison with the shadow of a Kerr BH ($\tilde{\xi}=0$, black dashed contour) is done.}
\label{F4}
\end{figure}
\begin{figure}[b!]
 \includegraphics[width=.44\linewidth]{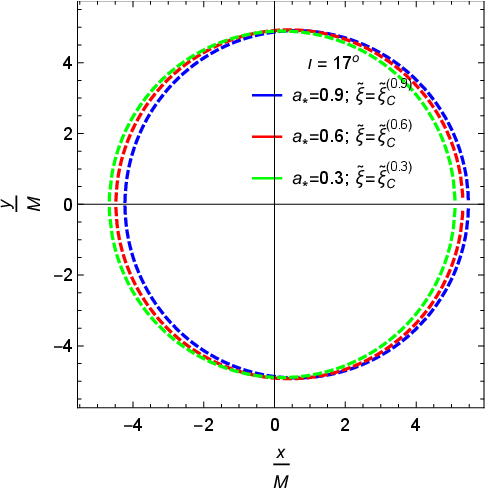}
 \includegraphics[width=.44\linewidth]{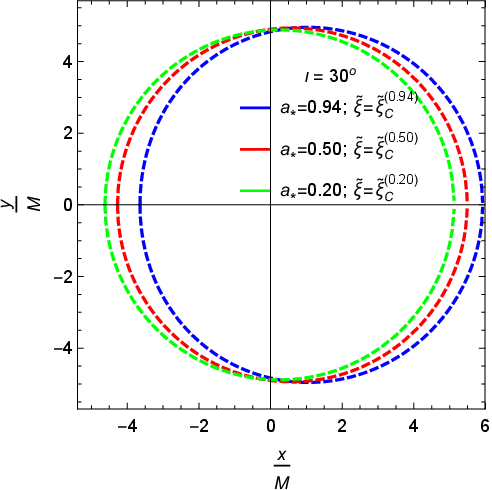}
\caption{Dependence of the shadow of the RGI rotating BH on the observer inclination angle. Left panel: 
$\iota=17^{\circ}$ and different values of $(a_{\star}, \tilde{\xi}^{(a_{\star})}_c)$. Right panel: The same as in the left panel but for $\iota=30^{\circ}$.}
\label{F5}
\end{figure}
From Eqs.~(\ref{eq41}) and (\ref{eq42}) it follows that the effect of the spin on the shadow contour is more pronounced for an observer inclination angle $\iota=\pi/2$. For this value of $\iota$, Fig.~\ref{F3} shows the shadow contour for selected values of $a_{\star}$ when $\tilde{\xi}=\tilde{\xi}_c$ for each one of these values. We see that for $\iota=\pi/2$ and for high values of the spin ($a/M \gtrsim 0.8$), just at the left end of the contour, given by one of the solutions of $y(r)=0$, appears a kind of cusp-like effect or ``point pressure deformation"  from the right that is a smooth dent for spins $a_{\star}\lesssim 0.8$ but increases as the spin goes to greater values. The left end of the shadow corresponds to photons in prograde orbits while the right end corresponds to photons in retrograde orbits. Consequently, the frame-dragging effect is stronger for prograde photons which leads to a greater distortion of the shadow of the RGI rotating BH than for the shadow of a classical Kerr BH. This is in line with the discussion in the previous section where we show that high values of the BH spin and high values of the parameter $\tilde{\xi}$ lead to a more compact horizon and, therefore, to prograde photons traveling through a region of higher curvature.\\

\noindent
Fig.~\ref{F4} shows the same as Fig.~\ref{F3} but in three separate panels in order to make evident two additional shadow features: (i) that the left cusp effect appears only in the immediate vicinity of $\tilde{\xi}=\tilde{\xi}_c$ (see the left panel), and (ii) that the area enclosed by the shadow contour of the RGI rotating BH is always smaller that the one enclosed by the shadow of a Kerr BH ($\tilde{\xi}=0$) which is shown as the black-dashed contour in each one of the three panels. Fig.~\ref{F5} reveals that the cusp effect is not detectable for observer inclination angles $\iota$ different from $\pi/2$, and that the increase in the spin value is accompanied by a displacement of the shadow to the right along the $x$-axis due also to frame-dragging.\\

\noindent
Interestingly, the same cusp effect that we notice is reported in Refs.~\cite{r27,r29}. There the authors succeed in obtaining a rotating regular metric in AS by using invariant scalars (as the Kretschmann scalar $K$) as the identification criterion with the energy scale $k$ in the form $k^2=\alpha K^{1/2}$ (where $\alpha$ is a dimensionless constant of order one), instead of the energy scale identification with the proper distance $k \sim 1/d$ discussed in Sec.~2. This seems to mean that for any RGI regular rotating BH, not only the presence of a smooth dent and a smaller shadow area are robust, but so is the presence of the cusp effect for $\iota=\pi/2$, for high values of $a_{\star}$ and for $\tilde{\xi} \approx \tilde{\xi}_c$. Obviously, as pointed out in \cite{r27}, if the quantum parameter $\xi$ is strictly tied to the Planck scale $\xi \sim m^{-4}_p$, as is done in AS, this effect is too tiny to be detectable with the current observational facilities (sew the following section).

\section{\label{sec:sec5}Shadow observables and comparison with the EHT data}

In this section we aim to constraint the quantum parameter $\tilde{\xi}$ using the recently EHT observations of the shadow images of the supermassive black holes M87$^*$ and Sgr A$^*$ \cite{r24,r25}.\\

\noindent
To characterize the shape of the shadow, several authors have proposed different sets of distortion parameters that can be measured from the observed EHT shadow images \cite{r43,r44,r45,r46,r47,r48,r49,r50}. Here we use the following three parameters:\\
 
(i) The areal radius $r_A$, which quantifies the size of the shadow and is defined by
\begin{equation}\label{eq43}
r_A=\sqrt{\frac{{\cal A}}{\pi}},
\end{equation}
where ${\cal A}$ is the area enclosed by the contour of the shadow
\begin{equation}\label{eq44}
{\cal A}=2\int y(r)dx(r)=2\int_{r_p^-}^{r_p^+} y(r)\frac{dx(r)}{dr} dr,
\end{equation}
and where $r_p^-$ and $r_p^+$ are the radii of photons, at unstable circular orbits, that co-rotate and counter-rotate, respectively, with the BH. These are the real positive solutions of $y^2(r)=0$.\\

\begin{figure}[t!]
 \includegraphics[width=.40\linewidth]{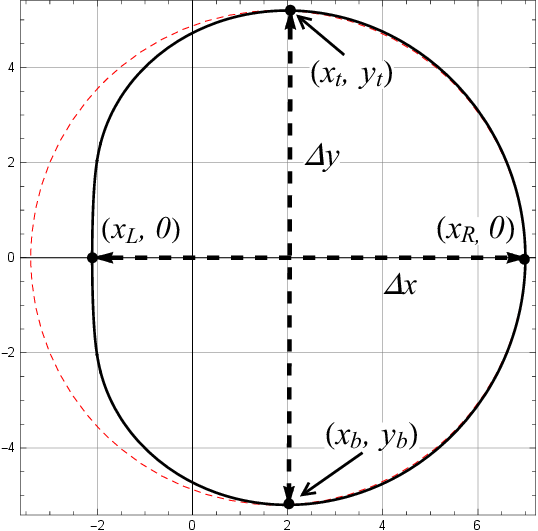}
\caption{The $(x,y)$ coordinates of the apparent shadow shape (solid black) of a rotating BH used to define the distortion parameters (for details see the text). The red-dashed curve is the so-called fitting circle which is a circle passing through the three points $(x_y,y_t)$, $(x_R,0)$ and $(x_b,y_b)$.}
\label{F6}
\end{figure}

(ii) The oblateness ${\cal D}$, which quantifies the deformation of the shadow, defined as
\begin{equation}\label{eq45}
{\cal D}=\frac{\Delta y}{\Delta x}=\frac{y_t-y_b}{x_R-x_L}=\frac{2y_t}{x_R-x_L},
\end{equation}
where the subscripts $L$, $R$ stand for the left and right ends of the shadow silhouette, and $t$, $b$ stand for the top and bottom ends. The factor 2 in Eq.~(\ref{eq44}) and the equality $y_b=-y_t$ in Eq.~(\ref{eq45}) are due to the reflection symmetry of the shadow respect to the $x$-axis as shown in Fig.~\ref{F6}.\\

(iii) The fractional deviation parameter $\delta_{sh}$ of the shadow diameter from that of a Schwarzschild BH, given by
\begin{equation}\label{eq46}
\delta_{sh}=\frac{R_{sh}}{3\sqrt{3}}-1,
\end{equation}
where $R_{sh}$ is the average radius of the shadow
\begin{equation}\label{eq47}
R_{sh}=\sqrt{\frac{1}{2\pi}\int_0^{2\pi}\left[(x-x_c)^2+(y-y_c)^2\right]d\phi},
\end{equation}
with
\begin{equation}\label{eq48}
\phi =\tan ^{-1}\left(\frac{y}{x-x_c}\right),
\end{equation}
and $(x_c,y_c=0)$ is the geometric center of the shadow with $x_c$ calculated from $x_c=(x_R-x_L)/2$.

\noindent
The left $x_L$ and right $x_R$ points are the celestial coordinate $x(r)$ evaluated at the real positive solutions of $y^2=0$, while the upper point $y_t$ of the shadow is obtained from the condition $dy/dx=0$, that is
\begin{equation}\label{eq49}
(1+{\cal M}^{\prime})r^3-3{\cal M}r^2+a^2(1-s^2)(1+{\cal M}^{\prime})r+a^2(1-s^2){\cal M} =0.
\end{equation}
where $s\equiv \sin{\iota}$.\\
\subsection{\label{sec:sec5.1}Constraints from the M87$^*$ data}
\begin{figure}[t!]
 \includegraphics[width=.44\linewidth]{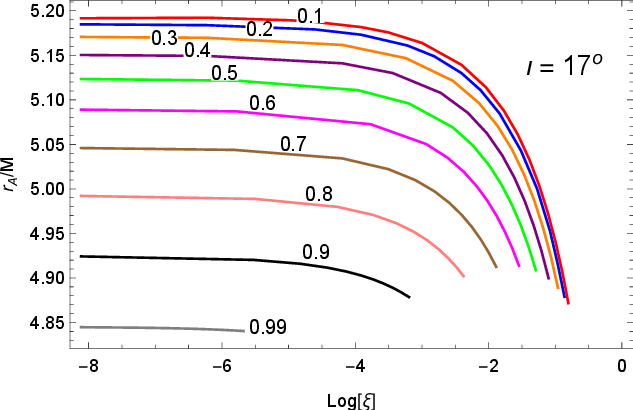}
 \includegraphics[width=.44\linewidth]{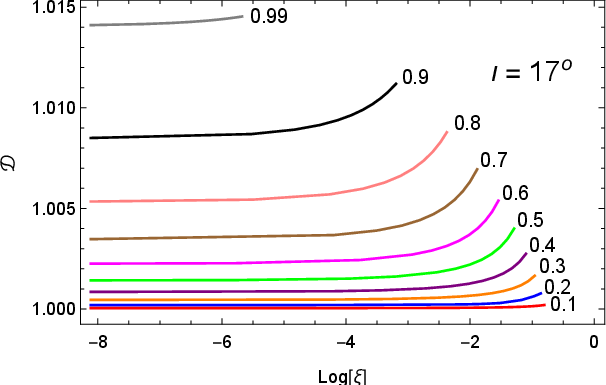}
\caption{Left panel: the areal radius per unit mass $r_{A}/M$ vs $\log(\tilde{\xi})$ of the RGI rotating BH for $\iota=17^{\circ}$ and for values of $a_{\star}$ in the range $0.1<a_{\star}<0.99$. Right panel: the oblateness ${\cal D}$ vs $\log(\tilde{\xi})$ for the same value of $\iota$ and for the same range of values of $a_{\star}$.}
\label{F7}
\end{figure}
The 2017 measurements of the shadow images of M87$^*$, obtained by EHT collaboration, determined the areal radius and the oblateness of this BH, respectively, in the ranges \cite{r24}
\begin{equation}\label{eq50}
4.31 M\leq r_A \leq 6.08,
\end{equation}
and
\begin{equation}\label{eq51}
1 \leq {\cal D} \leq 4/3,
\end{equation}
For the value of the observer inclination angle we take $\iota=17^{\circ}$ \cite{r52}.\\

\noindent
For the RGI rotating BH, the left panel of Fig.~\ref{F7} shows plots of $r_A/M$ vs. $\log \tilde{\xi}$ for different values of $a_{\star}$, while the right panel shows plots of ${\cal D}$ vs. $\log \tilde{\xi}$ for the same values of the spin parameter. The right end of each curve corresponds to the value of $\tilde{\xi}_c$ for each value of $a/M$ and the parameter $\tilde{\xi}$ runs in the rage $10^{-4}\leq \tilde{\xi}\leq \tilde{\xi}_c$. From the left panel we see that the quantum gravity effects render the size of the shadow smaller as $a_{\star}$ increases and that, for fixed value of $a_{\star}$, the size of the shadow decreases a $\tilde{\xi}$ grows. The change in size is more noticeable as  $\tilde{\xi}\rightarrow \tilde{\xi}_c$. The right panel shows that, in correspondence with the behavior of $r_A$, the shadow silhouette becomes more distorted as $a_{\star}$ becomes larger and, again, that the distortion is greater as $\tilde{\xi}\rightarrow \tilde{\xi}_c$. However, from these plots we also observe that both $r_A$ and ${\cal D}$ cover ranges entirely included within those established by the EHT collaboration, indeed: $4.840390\leq r_A/M \leq 5.191641$ and $1.000047 \leq {\cal D} \leq 1.014533$ for $0.1\leq a/M \leq 0.99$. This implies that, even tough the EHT measurements of $r_A$ and ${\cal D}$ are both satisfied, we can not determine constraints on the quantum parameter $\tilde{\xi}$ of the AS theory. In this sense, the present EHT measurements of M87$^*$ do not rule out the RGI regular rotating BH we are considering.\\
\subsection{\label{sec:sec5.1}Constraints from the Sgr A$^*$ data}
\begin{figure}[t!]
 \includegraphics[width=.44\linewidth]{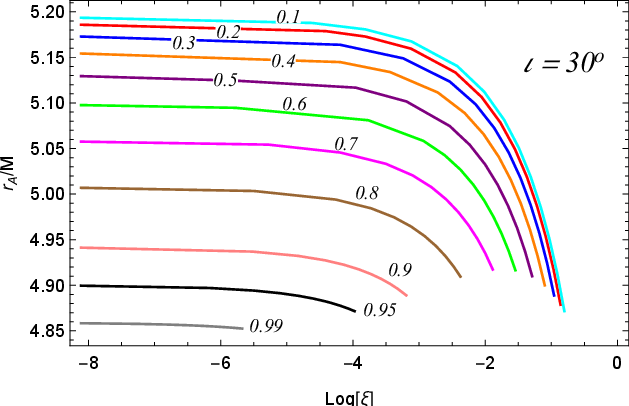}
 \includegraphics[width=.44\linewidth]{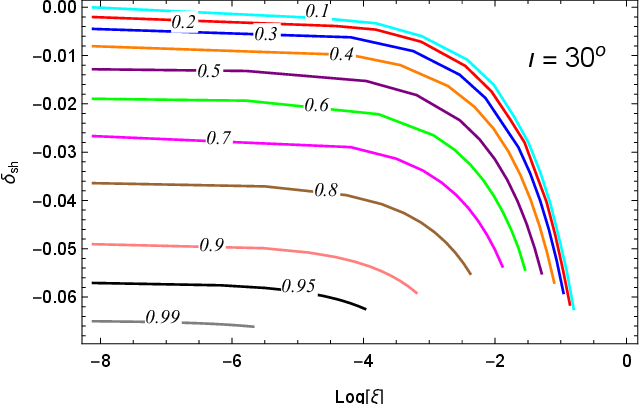}
\caption{Left panel: the areal radius per unit mass $r_{A}/M$ vs $\log(\tilde{\xi})$ of the RGI rotating BH for $\iota=30^{\circ}$ and for values of $a_{\star}$ in the range $0.1<a_{\star}<0.99$. Right panel: The fractional deviation parameter $\delta_{sh}$ for the same value of $\iota$ and for the same range of values of the spin parametr $a_{\star}$.}
\label{F8}
\end{figure}

To contrast our results with the EHT measurements for Sgr A$^*$ we use the data obtained by the Very Large Telescope
Interferometer (VLTI) and Keck observatories for the areal radius and for the fractional deviation parameter 
$\delta_{sh}$, respectively, as follows \cite{r25}
\begin{eqnarray}
4.5 M\leq r_A \leq 5.5,\quad \mathrm{(VLTI)}\\ \label{eq52}
4.3 M\leq r_A \leq 5.3, \quad \mathrm{(Keck)} \nonumber
\end{eqnarray}
\begin{eqnarray}
-0.17\leq \delta_{sh} \leq 0.01,\quad \mathrm{(VLTI)}\\ \label{eq53}
-0.14\leq \delta_{sh} \leq 0.05, \quad \mathrm{(Keck)} \nonumber
\end{eqnarray}
\noindent
The EHT collaboration has determined the observer inclination angle of Sgr A$^*$ to be $\iota < 50^{\circ}$ for two class of simulation models, one with $a_{\star}=0.5$ and the other with $a_{\star}=0.94$, that fully pass a set of EHT constraints for all the general relativistic magnetohydrodynamic (GRMHD) simulations, including the variability constraint. However, in order to fix a value of $\iota$ for our calculations, we take $\iota=30^{\circ}$ which is the value for the same two class of models passing all the EHT constraints but the variability constraints \cite{r25}. Our results for $r_A/M$ and $\delta_{sh}$ are shown in the left and right panels of Fig.~\ref{F8}, respectively, when $10^{-4}\leq \tilde{\xi}\leq \tilde{\xi}_c$ and for different values of $a/M$. First we see that the set of curves for $\delta_{sh}$ are almost a rigid displacement along the $y$-axis of the set of curves for $r_A$. This is because the values of $r_A$ and $R_{sh}$ calculated form the Eqs.~(\ref{eq43}) and (\ref{eq47}), respectively, are such that $r_A \approxeq R_{sh}$. The behavior of the size of the shadow ($r_A$) for different values of $a_{\star}$ and increasing values of $\tilde{\xi}$ is the same as the already discussed for the M87$^*$ BH. The absolute value of fractional deviation parameter $\delta_{sh}$ increases with increasing values of $a_{\star}$, with the increasing being more conspicuous when $\tilde{\xi}\rightarrow \tilde{\xi}_c$. The values of the areal radius run in the range $4.852527\leq r_A/M \leq 5.193447$ and the values of $\delta_{sh}$ cover the range $-0.0661306 \leq \delta_{sh} \leq -0.0000217$. Consequently, also in this case these ranges are subsets of the ones determined by the EHT collaboration, and we arrive again to the conclusion that the current EHT data for Sgr A$^*$ do not allow us to impose constraints on the parameter $\tilde{\xi}$ and that Sgr A$^*$ could be a RGI regular rotating BH.\\

\noindent
On the other hand, a recent measurement of the spin of the M87$^*$ BH from its observed twisted light, establishes that this BH rotates clockwise with a spin parameter $a_{\star}=0.9\pm 0.05$ \cite{r51}. Also, in a recent analysis of the spin of Sgr A$^*$ using the outflow method, results obtained with the preferred data sets were combined and indicate a practically identical spin parameter $a_{\star}=0.9\pm 0.06$ \cite{r52}. Then if, without pretending to be rigorous, we assume the central value $a_{\star}=0.9$ as the spin of both M87$^*$ and Sgr A$^*$, the quantum parameter $\tilde{\xi}$ is constrained to be in the range $0\leq \tilde{\xi} \leq \tilde{\xi}_c^{(0.9)}$, that is $0\leq \tilde{\xi} \leq 0.040967851$. This a very weak constraint if the value of the real physical parameter $\xi$ is conditioned by the Planck scale since, using for $M$ the mass of the BH Sgr A$^*$, we get $\xi = \tilde{\xi}^{(0.9)}M^4/m^4_{P} \lesssim 1.3\times 10^{177}$, where $m_{P}$ is the Planck mass. Clearly, the constraint is even weaker if we use the mass of M87$^*$.\\
\section{Summary and Conclusions}
The horizon-scale images of the supermassive BHs M87$^*$ and Sgr A$^*$ by the EHT collaboration, have opened an unprecedented window to test, in the strong-field regime, general relativity, alternative theories of gravity, as well as theories of quantum gravity. In this work we have studied quantum gravity effects on the shadow shape of a rotating BH in the framework of the AS theory for quantum gravity. We start from the RGI static regular exterior metric obtained in \cite{r21}, which results from the spherically symmetric collapse of dust. The collapse is regular due the to the antiscreening character of gravity at small distances in AS that leads to a ball of dust that never ends in a singularity at any finite time. Next, we have used the generalized NJ algorithm to construct the RGI rotating metric. This metric differs from the Kerr solution by the replacement of the constant BH mass by a Misner-Sharp mass function ${\cal M}(r,\tilde{\xi})$ depending only on the radial coordinate and on a dimensionless free parameter $\tilde{\xi}$ which captures the quantum gravity effects.\\

\noindent
The study of the causal structure of the RGI rotating BH space-time has revealed that, associated to each value of the spin parameter $a_{\star}$ ($0<a_{\star}\leq 1$), there is a critical value $\tilde{\xi}^{(a_{\star})}_c$ such that for $0<\tilde{\xi}<\tilde{\xi}_c$ there are two horizons, one outer horizon and one inner Cauchy horizon; for $\tilde{\xi}=\tilde{\xi}_c$ the two horizons coincide, and for $\tilde{\xi}>\tilde{\xi}_c$ a horizonless scalar remnant arises. It is important to note that for growing values of the BH spin and growing values of $\tilde{\xi}$ ($\tilde{\xi}\leq\tilde{\xi}_c$) the horizon shrinks, that is, becomes more compact. We have also argued that, being the outcome of a ball of dust that never reaches the singularity, the metric describing the RGI rotating BH is also free of singularities.\\

\noindent
When analyzing the shadow cast by the RGI regular rotating BH, we have found the following main features: (i) the area enclosed by its shadow contour is always smaller that the area enclosed by the shadow of a Kerr BH 
($\tilde{\xi}=0$), (ii) {\it only} for an observer inclination angle $\iota=\pi/2$ and for values of the spin parameter $a_{\star} \gtrsim 0.8$ and $\tilde{\xi} \approx \tilde{\xi}_c$, a kind of cusp-like effect appears which increases as $a_{\star}$ increases (see Fig.~\ref{F3}). Since the left end of the shadow corresponds to photons in prograde orbits, the frame-dragging effect is stronger for these photons than in the case the clasical Kerr BH. This is in accordance with the fact that high values of the BH spin and high values of the parameter $\tilde{\xi}$ lead to a more compact horizon. The cusp effect bears great similarity to the found in \cite{r27,r29} and lead us to argue that the two referred features are robust for shadows cast by RGI regular rotating BHs.\\

\noindent
To confront our findings with the EHT images of M87$^*$ and Sgr A$^*$, we have used the following shadow observables: for comparison with the M87$^*$ data we use the areal radius $r_A$, that inform us on the size of the shadow, and the oblateness ${\cal D}$, which quantifies the deformation in the shadow, and for comparison with the Sgr A$^*$ data we use the areal radius $r_A$ and the fractional deviation parameter $\delta_{sh}$ of the shadow diameter with respect to the shadow of a Schwarzschild BH ($\delta_{sh,S}=0$). Our results for the RGI rotating BH are shown in Figs.~\ref{F7} and ~\ref{F8}. We have found that the range of values of all these observables are proper subsets (in the mathematical sense) of the ranges established by the EHT collaboration for them. Our numerical results for $r_A$ are essentially the same as the reported in \cite{r53}. We then are forced to conclude that the current EHT data for M87$^*$ and Sgr A$^*$ do not allow us to impose constraints on the quantum parameter $\tilde{\xi}$ and, for the same reason, that the RGI regular rotating BH is not ruled out by these data. We note, however that, as discussed above, if we take for $a_{\star}$ the central value of the recently established spin parameters: $a_{\star}=0.9\pm 0.05$ for M87$^*$ \cite{r51}, and $a_{\star}=0.9\pm 0.06$ for Sgr A$^*$ \cite{r52}, and we use the mass of the BH Sgr A$^*$, we can impose on the physical parameter $\xi$ the very weak constraint $\xi = \tilde{\xi}^{(0.9)}M^4/m^4_{P} \lesssim 1.3\times 10^{177}$.\\
 
\noindent
It is to be expected that future improvement of the image fidelity of the EHT array will allow to better constrain the image morphology of the BH shadows and, hopefully, to identify quantum gravity effects from them.

\section*{ACKNOWLEDGEMENTS}
The author is thankful for useful and insightful discussions with Oleksandr Stashko. The author acknowledge financial support from Universidad Nacional de Colombia.

\end{document}